\documentclass[a4paper,12pt]{article}

\usepackage{a4wide}
\usepackage{graphicx}
\usepackage[utf8]{inputenc}
\usepackage{authblk}
\usepackage{xcolor}

\newcommand{\uniboaffil}{\small Department of Computer Science and Engineering, Campus of Cesena, Universit{\`a} di Bologna}
\newcommand{\ecltaffil}{\small European Centre for Living Technology, Venezia, Italy}
\newcommand{\isbaffil}{\small Institute for Systems Biology, Seattle, USA}

\begin{document}

\author[1,2]{Andrea~Roli}
\author[3]{Stuart~A.~Kauffman}

\affil[1]{\uniboaffil}
\affil[2]{\ecltaffil}
\affil[3]{\isbaffil}

\date{}

\title{The Hiatus Between Organism and Machine Evolution: Contrasting Mixed Microbial Communities with Robots}

\maketitle

\begin{abstract} 
Mixed microbial communities, usually composed of various bacterial and fungal species, are fundamental in a plethora of environments, from soil to human gut and skin. Their evolution is a paradigmatic example of intertwined dynamics, where not just the relations among species plays a role, but also the opportunities---and possible harms---that each species presents to the others. These opportunities are in fact \textit{affordances}, which can be seized by heritable variation and selection. In this paper, starting from a systemic viewpoint of mixed microbial communities, we focus on the pivotal role of affordances in evolution and we contrast it to the artificial evolution of programs and robots. We maintain that the two realms are neatly separated, in that natural evolution proceeds by extending the space of its possibilities in a completely open way, while the latter is inherently limited by the algorithmic framework it is defined. This discrepancy characterises also an envisioned setting in which robots evolve in the physical world. We present arguments supporting our claim and we propose an experimental setting for assessing our statements. Rather than just discussing the limitations of the artificial evolution of machines, the aim of this contribution is to emphasize the tremendous potential of the evolution of the biosphere, beautifully represented by the evolution of communities of microbes.
\end{abstract}

\section{Introduction}

The debate between Georges L{\'e}opold Dagobert, Baron Cuvier, and {\'E}tienne Geoffroy Saint Hilaire regarding the causal relation between form and function of animal parts~\cite{avery2021information} highlights the central role of shapes and functions in the evolution of organisms. Geoffroy wrote that ``Animals have no habits but those that result from the structure of their organs: if the latter varies, there vary in the same manner all their springs of action, all their facilities, and all their actions.''~\footnote{Text quoted in~\cite{avery2021information}, page 12.}, and he maintained that the same structure in different organisms may express different functions. Geoffroy's perspective emphasizes the role of the environment and the actions that the ``organ'' enables the organism in that environment. Two centuries later, the central role of the environment and the possibilities offered by organism ``parts'' in evolution is certainly acknowledged, but probably not as widely as it deserves. Foremost in this respect is, we believe, the notion of \textit{affordances}~\cite{walsh2015organisms}. Here we refer to the evolutionary viewpoint of affordances, i.e. those opportunities typically seized by heritable variation and natural selection. A notable example is that of flight feathers, which initially evolved for thermal insulation or as bristles, but subsequently has been used for enabling flight~\cite{prum2002evolutionary}, or the swim bladder~\cite{Kauffman2016-humanity}. Once a new function for an organism ``part'' is established, it becomes available for further uses or as a component in the descendants.

Mixed microbial communities (MMCs) are characterized by a wide spectrum of phenomena in which the opportunities offered by the environment---communities themselves included---are vital for evolutionary adaptation~\cite{gorter2020understanding}. Furthermore, MMCs are a prominent illustration of the intertwined paths natural evolution traverses, acting at the same time across the whole organization of the system, from the molecular level to interaction among communities. We maintain that this property is typical of biological systems and cannot be achieved in the artificial evolution of computer programs and robots. Contrasting the natural evolution of MMCs with the artificial evolution of robots makes it possible to highlight the distinguishing generic properties of biological systems and understand the limits of artificial evolution of machines, as well as to appreciate the flourishing potential of the evolution in our biosphere. In addition, we propose an experimental setting for assessing the discrepancy between natural and artificial evolution.

\section{The role of affordances in evolution}

Most adaptations in evolution are \textit{affordances}, generally seized by heritable variation and natural selection.\footnote{Heritable variations include gene mutation and also epigenetic phenomena~\cite{jablonka2002changing}.}
Gibson~\cite{Gibson1966} used the term ``affordance'' to represent the fact the ``things'' perceived by organisms in an environment afford them possible ``actions''. These  ``things'' can be inanimate structures, such as rocks, or other organisms; by ``actions'' we mean activities pursued by the organism in its environment and functions performed by the organism to support its maintenance and ultimately its survival. For example, a hole in a rock affords a wasp a safe place to build a nest, or a shell---with suitable characteristics---affords a home for the hermit crab \textit{Pagurus bernhardus}.
Since the initial work by Gibson in ecological psychology, the notion of affordance has been elaborated in diverse fields, such as biosemiotics~\cite{campbell2019learning}, robotics~\cite{jamone2016} and philosophy~\cite{heras2019philosophy}.
In general, an affordance can be defined as ``a possible use of X by an organism to accomplish Y''. Here we focus on ``accomplishments'' that can occur by heritable variation and selection, as in the evolution of the heart and the loop of Henle~\cite{kauffman2021world}.
Crucially, affordances are not independent features of the environment~\cite{walsh2015organisms}, as a change can be neutral or not depending on the conditions of the organism, its ``goals'' and its repertoire of actions~\cite{roli2020emergence,roli2022organisms}. 

The history of evolution is rich of examples of adaptation emerged by co-opting the same organ for a new function, i.e. of so-called Darwinian preadaptations or exaptations~\cite{gould1982exaptation}. In this case, the organ affords a new use for the organism.
Typical examples are flight feathers, evolved for accomplishing functions such as thermal insulation and then co-opted for the new function of flight~\cite{prum2002evolutionary,persons2015bristles}, and lens crystallins, which originated first as enzymes~\cite{barve2013latent}. A prominent example, discussed in~\cite{Kauffman2016-humanity}, is the evolution of the swim bladder, which enables fishes to attain neutral buoyancy. Paleontologists conjecture that the swim bladder has arisen from the lungs of lung fish: This is precisely finding a new use for the same initial organ.

The evolutionary adaptation to new environmental conditions is often achieved by organisms' \textit{phenotypic plasticity}, whereby adaptation occurs without requiring a genetic variation; however, this phenotypic variant can be inherited if produced by epigenetic differences in gene expressions~\cite{thorson2017epigenetics}, or it can bias the subsequent selection of favourable genetic traits~\cite{West-Eberhard6543}.

A new use can be driven by random events and accidents, such as genetic mutations, or by a change in the environment. 
Changes in the environment consist in any alteration that happens in the portion of the environment that the organism can perceive and with which it can interact, including other organisms. This ``niche'' represents the \textit{umwelt} of the organism, according to von~Uexk{\"u}ll~\cite{Uexkull-foray-correct,kull1998semiosis}.
The biosemiotic viewpoint of affordances~\cite{kull2015evolution} is crucial here: a mutation or a change in the environment triggers the need for identifying affordances, and the ones useful for the organism can be seized by selection.

As the organism's \textit{umwelt} also includes other organisms, the systems composed of more than one species are a fabulous example of higher-order dynamics. For instance, the by-product or waste of a species might be exploited by another species able to take profit of this niche, i.e. the waste produced by an entity can be seen as an affordance by another entity.
In this perspective, also symbiosis can be interpreted as the result of combinations of seized affordances, whereby species interaction provides reciprocal and long-term coupling sustaining their life. Nature abounds in examples of symbiotic structures, spanning the whole spectrum of life beings~\cite{margulis2008symbiotic}. 
Ecosystems and in general systems composed of several species have an evolutive advantage when they can mutually offer each other complementary capabilities~\cite{corning2021vs}. This symbiotic cooperation also produces rich semiotic networks~\cite{sharov2016evolutionary}, which confer both robustness and plasticity by enabling multiple alternative signaling pathways and diverse mechanisms to act in the environment. 
These properties are typical of mixed microbial communities (MMCs), whose main characteristics are succinctly illustrated in the following section.

\section{A Systemic View of Mixed Microbial Communities}

Communities composed of several species of micro-organisms, such as bacteria and fungi, are pervasive in nature, from marine environments to the microbiome in human guts. These communities are currently of great interest and are studied both for understanding the general mechanisms of their dynamics and evolution~\cite{prosser2020conceptual}, and for their applications in human health and agriculture~\cite{eng2019microbial}. 

The dynamics underpinning their activity and their evolution have still to be fully understood~\cite{prosser2020putting}.
The dynamical properties of a community of microbes, and in particular its stability, robustness and adaptiveness, is the result of the networks of interactions among its composing organisms.
Some key interactions~\cite{ravikrishnan2018systems} sustaining their dynamics are \textit{mutualism}, where all organisms involved benefit from the interaction; \textit{commensalism}, where only one organism or species has a benefit; and \textit{parasitism}, where one organism has a benefit while harming the others.

The composition and the kind of interactions of MMCs change along evolution, taking profit of new possibilities offered by the colonies and adapting to possible environmental variations~\cite{castledine2020community,gorter2020understanding}. Different environmental conditions give rise to possibly different composition in the community, different interactions and different community proteomes~\cite{siggins2012exploring}. A case in point is the emergence and dynamical creation of new niches, favored also by horizontal gene transfer mechanisms~\cite{baquero2021origin}. From a cybernetic perspective~\cite{Ashby-design}, this evolutionary adaptation coincides with the creation of meaning, in that the system reorganizes and changes itself to respond to environmental changes, attributing sense to what is relevant to its survival~\cite{roli2020emergence,igamberdiev2021mathematics}.

The interactions among micro-organisms and communities not only take place at different levels, from the individual organism to the community of diverse species, but the levels themselves interact in a mixture of upward and downward causation~\cite{emmeche2000levels,noble2012theory,Kauffman2019beyond}. In the former, causal processes at the lower level produce a higher level phenomenon. Hence, the explanatory arrow points downward. While in downward causation a process or an entity has causal effects to lower level entities. Hence, in this case the explanatory arrow points upward.
A representative illustration of emergent dynamics across all the levels is that of \textit{metabolically cohesive microbial consortia}~\cite{pascual2020metabolically}, which are stable intermediate-level entities, composed of some species, emerging in MMCs. These structures are dynamically created as a response to environmental conditions and their structure can be modified if conditions change. Their evolutive advantage is attributed to a more efficient consumption of resources. Cohesive consortia are a paradigmatic example in our discussion, for they represent a prominent case in which affordances are identified and provide a reciprocal advantage for the entities involved in the consortium. Moreover, not only the cohesive consortia can be favored by selection, but the organisms composing them remain individually available for selection~\cite{kauffman2021third}. Therefore, the appearance of a cohesive consortium extends the \textit{adjacent possible} of the mixed community~\cite{Kauffman2016-humanity}. Affordances are in fact degrees of freedom of biological systems~\cite{Kauffman2019beyond}. A seized affordance produces a non-reversible effect in the biosphere as it extends the phase space of possibilities: There is something that is possible now, which was not possible before. The evolution of the biosphere is then moved by affordances and, astonishingly, no mathematics based on Set Theory---hence algorithmic---can be used to predict its evolution~\cite{kauffman2021world}. 

Finally, we observe that living systems are dynamically critical, poised between order and disorder~\cite{roli2018dynamical}. The reasons for this property are to be found in the capability of optimally balancing \textit{(i)} discrimination of inputs and reliability of actions---at the individual adaptation level---, and \textit{(ii)} adaptiveness and robustness against mutation---at the evolutionary level. Remarkably, repeated interactions among the entities of a system subject to evolution are likely to lead to systems composed of critical entities~\cite{kauffman2010coevolution,Hidalgo2014}. Although in the literature there are no results on criticality of MMCs,\footnote{To the best of our knowledge} we expect also these systems to exhibit this dynamical property.

\section{Artificial Evolution of Robots}
\label{sec:ER}

Artificial Life has a long record of studies concerning the artificial evolution of \textit{in silico} organisms.\footnote{See, e.g. the ALife conference series~\cite{ALife}.} The objectives of studies in evolution of artificial creatures are multiple and involve the investigation of the conditions for the emergence of target properties or capabilities, the exploration of possible evolutionary scenarios, and system design applications, just to cite a few. 
The fundamental machinery of artificial evolution consists in iteratively applying variation and selection---and possibly other genetic operators---on individuals encoded in a formal representation (e.g. a LISP program), where selection is based on an objective function---the \textit{fitness function}---that evaluates the quality of the individuals with respect to a given merit factor. The fitness function can also be implicitly defined by designing an environment in which the digital creatures compete for resources.

We have previously addressed elsewhere the main issues in artificial evolution of organisms~\cite{roli2020emergence} and the limits of Artificial General Intelligence~\cite{roli2022organisms}, emphasizing the crucial role played by affordances and the creation of meaning. Here we just summarize the main points and refer the interested reader to the specific papers~\cite{roli2020emergence,kauffman2021world,roli2022organisms}, and the references they contain. The cardinal statement is that since algorithmic procedures can only identify a limited set of affordances, artificial evolution---inasmuch based on Universal Turing Machines (UTMs)---is inherently limited compared to the natural one.
A computer simulation of an evolutionary process, such as the famous Tierra~\cite{ray_evolution_1992} in which digital organisms are evolved in a digital world, can only produce results inside the framework of the data structures and the operations that have been defined by the designer. For example, the interactions among digital organisms are either predefined or the result of predefined compositional rules. Such a simulation can surely produce outcomes that surprise the human observers, but it cannot produce anything outside the ``ontology'' is has been provided. The space of possibilities in a computer program is like a LEGO\textsuperscript{\scriptsize{\textcircled{\tiny{R}}}} bricks world, where the basic components and the rules for combining them are already given. Magnificent constructions can be built, but none of them can produce pieces combined in an illegal way, e.g. at an angle of 27 degrees, neither can they contain pieces of arbitrary shape. Only if we add tape and a cutter we are able to do this, i.e. only if we arbitrarily break the rules of the game. Note that there are no boundaries to the ways we can break the rules of the game. In the evolution of the biosphere these rules are constantly changed, and in an unpredictable way as they depend upon affordances. An objection might be raised by observing that living organisms are essentially made of elementary particles, which are somehow given along with the interactions among them; therefore, one could in principle define a phase space in which trajectories of evolution can be depicted and possibly predicted with an error that can be reckoned. This is true for a physical space, in perfect agreement with the Newtonian Paradigm,\footnote{This holds also for quantum mechanics, even if in probability.} where the degrees of freedom and so the phase space are known in advance, but it falls short for the biosphere, where affordances are the degrees of freedom, which appear unpredictably. In this space, no error measure can be defined~\cite{longo2013perspectives}, so no sound prediction can be made, but just educated or wild and haphazard guesses.

In the case of evolution of communities composed of different species, the limits of simulated evolution are even stronger. An experiment can be conceived to assess the discrepancy between evolution of natural and artificial MMCs and and will be discussed in Section~\ref{sec:experiments}.

One might ask if robots subject to evolution can exploit their property of being \textit{embodied}~\cite{pfeifer2006How} and so reduce or even clear this discrepancy. Robotics and evolution are combined in Evolutionary Robotics (ER), which is a branch of Artificial Life and Artificial Intelligence, in which robots are designed by means of an artificial evolution process~\cite{Nolfi-and-Floreano2000,Bongard-ACM2013}.
The typical experimental setting in ER consists in artificially evolving a population of robot controllers (e.g. an artificial neural network) such that a robot with desired performance at a given task is produced. In some notable studies, also the morphology of robots is evolved~\cite{pollack2003computer,blackiston2021cellular}. Usually, owing to practical constraints, robot performance is evaluated in simulation. 
Notable achievements in ER have been attained, especially showing that artificial evolution can produce robots with high performing behavior as well as capable of showing novel solutions---somehow surprising---to the designers~\cite{Bongard-ACM2013,Doncieux2015}.

As long as robots are simulated, i.e. a model of them and of the environment is run in a computer program, the limits of the ``LEGO\textsuperscript{\scriptsize{\textcircled{\tiny{R}}}} bricks world'' still hold.
However, a recent remarkable work presents an automated system that evolve robots in the physical world, thus exploiting the physical properties of robots and their environment~\cite{hale2019robot}. In that work, robots are assembled from basic building blocks by an automatic factory that follows given rules. Then, the performance of the robot is assessed in the physical world.
If we project this idea to a context in which the fitness function is not defined by the designer, but rather implicitly expressed as the survival of the robot, we ask whether this might make it feasible to achieve a fully-fledged evolution of artificial entities.
Furthermore, systems undergoing coevolution have been proven to attain high performance and the evolutionary process can explore the space of possible configurations more efficiently~\cite{koza1991genetic} and can exhibit generic properties typical of living systems, such as criticality~\cite{kauffman1996home,munoz2018colloquium}. If we add to coevolution mechanisms the possibility of evolving colonies of heterogeneous robots---in the same way as MMCs evolve---we might believe to be very close to achieving true \textit{open-ended evolution} of embodied machines.
We maintain that, despite these thrilling and perhaps threatening plausible scenarios, there is a hiatus between the evolution of organisms and the evolution of robots that cannot be bridged (at least in practice). We elaborate our arguments in the next section.

\section{Contrasting the evolution of living beings with the artificial evolution of robots}
\label{sec:discussion}

Our argument is based on five pillars:

\vspace{1ex}
\noindent
\textit{(i)} Living organisms evolve together with their \textit{sensors} and \textit{actuators}---being these terms used in analogy to robots. Sensors are the means by which organisms perceive what is relevant to them in the world, and, by virtue of this, they attribute meanings to the world. Actuators are the tools that enable organisms to act purposively. As Cariani beautifully depicts: ``Sensors and effectors are the crucial points at which a real world situation is encoded into a symbolic representation and at which an action-decision is transformed from symbolic representation into physical action. It is only by virtue of actual connection to the world via sensors and effectors that symbolic representations become semantically grounded.''~\cite{cariani1998towards}. The main sensors and actuators of micro-organisms are proteins. As long as machines cannot create their own sensors and actuators, this crucial stage cannot be met because they will only be able to perceive what has been connoted with meaning by the designers, and perform actions affecting what the designers have identified as relevant in the environment. Therefore, a question arises as to what extent robots will be able to create their own sensors and actuators. This creation can be fully achieved only if robots will be capable of \textit{(a)} building artefacts at the level nature can build, and \textit{(b)} generating their own goals. The first capability is unlikely to be achieved but maybe feasible if robots are built of biological material. The second property is needed because only what is relevant to the organism makes it possible to see and identify affordances. This requires \textit{organismal agency}, which we illustrate in the second pillar.

\vspace{1ex}
\noindent
\textit{(ii)} Truly open-ended evolution can be achieved only by \textit{organismal agency}~\cite{jaeger_fourth_2022}, which grants ``co-emergent dialectic between organisms' goals, actions, and affordances''~\cite{roli2022organisms}, and makes it possible for evolution to transcend its predefined phase space. Organismal agency does not just characterize humans and animals in general, but also bacteria~\cite{fulda2017natural}. Therefore, MMCs are the result of co-evolution of colonies of micro-organisms to which we can attribute the capability of seeing the world, evaluating what is ``good or bad for me'', and acting~\cite{peil2014emotion}.

\vspace{1ex}
\noindent
\textit{(iii)} If the robot control system is a computational unit implementing a UTM, then robots cannot be creative~\cite{alexander2021living,roli2022organisms}. Nevertheless, affordances can  also be seen by robots just by random accidents~\cite{roli2020emergence}. For example, a collision against a hard object could produce a dent in the robot's chassis; this dent could serendipitously be useful to detect places for recharging, so this change in the morphology could turn out to be an advantage for the robot. This property can be included in the evolutionary process only if it can be somehow modeled as a heritable trait or the robot can give it a name and make explicit use of it. Both the conditions have to face the symbol grounding problem~\cite{Harnad1990}, which again requires organismal agency to be tackled~\cite{roli2022organisms}. In addition, finding affordances by pure chance requires an overwhelmingly long time scale. 

\vspace{1ex}
\noindent
\textit{(iv)} The ``evolutionary machinery'' must be subject to evolution, too, because in nature also genetic operators evolve together with the organisms---see, for example, horizontal gene transfer in bacteria. Therefore, these mechanisms cannot be based on a predefined set of formal rules or instructions, but it should be embedded in the very same evolutionary setting of the organisms. Furthermore, the ``code'' describing the construction of robots should be evolvable as well and, again, it must not be based on a formal language, which would introduce the limitations already illustrated in Section~\ref{sec:ER}.

\vspace{1ex}
\noindent
\textit{(v)} Current robots have very few levels of organization and they can be easily articulated according to their parts and functions. This is not the case for micro-organisms, and living systems in general~\cite{kauffman1970articulation}. This situation is even more striking in MMCs, where we also observe higher levels of organization. Therefore, the interactions among parts in robots and among robots should be left completely open to evolve, which means that they have not to be defined in a formal language with predefined semantics.

\section{Assessing the discrepancy between evolution \mbox{\textit{in silico}} and in nature}
\label{sec:experiments}

The evolution of MMCs can now be experimentally studied in a feasible experimental setting: $N$ samples from a MMC are taken and are cultivated independently in $N$ separated and identical bioreactors~\cite{auchtung2015cultivation,guzman2018using}. Their features, such as diversity of species, composition, community metabolome, genome and proteome, can be studied over time in each of the $N$ samples as the members of the community evolve by changing their relative abundances and accumulating mutations by means of current available technologies~\cite{chandran2020microbial,siggins2012exploring,riesenfeld2004metagenomics,metch2018metagenomic,feng2018metagenomic,chacon2020genomic}. This produces a time dependent fingerprint for each of the $N$ colonies, which can then be compared. The metabolome, proteome, genomic, and metagenomic analysis of a MMC provides a rich source of information on the evolving organization of the colony, as it makes it possible to identify the species, their relations, the emergence of novel functions of existing proteins, and the time course of the accumulated mutations. A quantitative analysis of metabolome, proteome and genomic diversity across the $N$ colonies provides an estimation of the diversity of evolutionary paths taken by the MMCs. The same experiment allows us to assess if the $N$ initial samples of the MMC evolve to the same, similar or very distinct final compositions. Similar studies allow us to ask if initially divergent communities evolve to the same or different final communities.

If an analogous experiment could be designed also for the evolution of artificial organisms, then an assessment of the discrepancy between evolution in MMCs and in robot groups could be achieved.

A technologically currently feasible experiment can be conceived regarding the evolution of communities of agents in simulation.
This experiment consists in evolving populations of Boolean networks~\cite{kauffman1969metabolic,Kau1993}. The Boolean networks (BNs) can interact by connecting some of their nodes: the current value of a network node can be overridden by the node value of another network. These connections may be defined via a kind of genetic code, which also comprises topology and functions of the network. The selective pressure can be defined in terms of values assumed in time by BN \textit{essential variables}~\cite{Ashby-design}. In the initial condition the colony should be composed of heterogeneous communities of homogeneous networks, i.e. networks with the same topology and Boolean functions---and all other components of the ``genetic code'' which describes them. An assessment of their ``proteome'' can be done by clustering the genetic descriptions of the BNs by means of the Normalized Compression Distance~\cite{cilibrasi2005clustering}, which can be applied to any data structure. The comparison of different replicas of the experiment makes it possible to quantitatively assess the diversity of the final ``artificial proteome'', plus several other properties such as communities and their interactions, and generic dynamical properties such as criticality.

Moving towards embodied robots we might envision an experiment involving physical robots controlled by BNs. An evaluation of the robot in the physical environment is time demanding, but in principle it can be done. Nevertheless, further limitations are imposed by the morphology of the robots, and their sensors and actuators, which should be evolvable. We believe that such an experiment has currently too many technological obstacles to be realized. However, the evolution of robots made of biological material is highly promising and could be the way to design an experiment of ``embodied machines'', analog to that of communities of BNs.

\section{Conclusion}

MMCs are a wonderful example of systems exhibiting a rich dynamics of intertwined levels, balancing robustness and adaptiveness, which manifests itself in a kaleidoscope of different functions and capabilities. A wealth of examples of MMCs can be observed in nature, and, due to fast genetic mechanisms, their evolution can be studied in reasonably short amounts of time. A systemic view of MMCs makes it possible to compare them with communities of artificial machines and contrast the evolution of MMCs with the (envisaged) evolution of communities of robots. The aim of this comparison is to provide a support to the thesis that living beings and their evolution are inherently and inexorably different than machines based on UTMs and their artificial evolution: The former are capable of making sense of their world and be creative, while the latter cannot create information and are only able to achieve some kind of combinatorial creativity or innovation. By elucidating this difference we are not arguing that computational methods cannot be used to study MMCs \textit{tout court}, but rather than they can be used to address generic properties, and we propose to experimentally assess the discrepancy between the two realms.
If we accept this hiatus, then have to acknowledge that the evolution of robots based on UTMs will never be capable of creativity and true innovation. Maybe this is possible with robots made of biological material, but the question arises as whether we really aim at this technology.

\section*{Acknowledgements}
We thank Jan Dijksterhuis for fruitful discussions about soil communities.


\begin{thebibliography}{10}

\bibitem{ALife}
{International Society for Artificial Life}.
\newblock https://alife.org.
\newblock Accessed: 2022-06-02.

\bibitem{alexander2021living}
V.N. Alexander, J.A. Bacigalupi, and {\`O}.C. Garcia.
\newblock Living systems are smarter bots: Slime mold semiosis versus {AI}
  symbol manipulation.
\newblock {\em Biosystems}, 206:104430, 2021.

\bibitem{Ashby-design}
W.R. Ashby.
\newblock {\em Design for a brain: The origin of adaptive behaviour}.
\newblock Second edition, 1954.

\bibitem{auchtung2015cultivation}
J.M. Auchtung, C.D. Robinson, and R.A. Britton.
\newblock Cultivation of stable, reproducible microbial communities from
  different fecal donors using minibioreactor arrays ({MBRAs}).
\newblock {\em Microbiome}, 3(1):1--15, 2015.

\bibitem{avery2021information}
J.S. Avery.
\newblock {\em Information theory and evolution}.
\newblock World Scientific, Singapore, third edition, 2021.

\bibitem{baquero2021origin}
F.~Baquero, T.M. Coque, J.C. Gal{\'a}n, and J.L. Martinez.
\newblock The origin of niches and species in the bacterial world.
\newblock {\em Frontiers in microbiology}, 12:657986, 2021.

\bibitem{barve2013latent}
A.~Barve and A.~Wagner.
\newblock A latent capacity for evolutionary innovation through exaptation in
  metabolic systems.
\newblock {\em Nature}, 500(7461):203--206, 2013.

\bibitem{blackiston2021cellular}
D.~Blackiston, E.~Lederer, S.~Kriegman, S.~Garnier, J.~Bongard, and M.~Levin.
\newblock A cellular platform for the development of synthetic living machines.
\newblock {\em Science Robotics}, 6(52):eabf1571, 2021.

\bibitem{Bongard-ACM2013}
J.C. Bongard.
\newblock Evolutionary robotics.
\newblock {\em Communications of the ACM}, 56(8):74--83, 2013.

\bibitem{campbell2019learning}
C.~Campbell, A.~Olteanu, and K.~Kull.
\newblock Learning and knowing as semiosis: Extending the conceptual apparatus
  of semiotics.
\newblock {\em Sign Systems Studies}, 47(3/4):352--381, 2019.

\bibitem{cariani1998towards}
P.~Cariani.
\newblock Towards an evolutionary semiotics: the emergence of new
  sign-functions in organisms and devices.
\newblock In {\em Evolutionary systems}, pages 359--376. Springer, 1998.

\bibitem{castledine2020community}
M.~Castledine, P.~Sierocinski, D.~Padfield, and A.~Buckling.
\newblock Community coalescence: an eco-evolutionary perspective.
\newblock {\em Philosophical Transactions of the Royal Society~B},
  375(1798):20190252, 2020.

\bibitem{chacon2020genomic}
K.~Chac{\'o}n-Vargas, J.~Torres, M.~Giles-G{\'o}mez, A.~Escalante, and J.G.
  Gibbons.
\newblock Genomic profiling of bacterial and fungal communities and their
  predictive functionality during pulque fermentation by whole-genome shotgun
  sequencing.
\newblock {\em Scientific reports}, 10(1):1--13, 2020.

\bibitem{chandran2020microbial}
H.~Chandran, M.~Meena, and K.~Sharma.
\newblock Microbial biodiversity and bioremediation assessment through omics
  approaches.
\newblock {\em Frontiers in Environmental Chemistry}, 1:570326, 2020.

\bibitem{cilibrasi2005clustering}
R.~Cilibrasi and P.M.B. Vit{\'a}nyi.
\newblock Clustering by compression.
\newblock {\em IEEE Transactions on Information theory}, 51(4):1523--1545,
  2005.

\bibitem{corning2021vs}
P.A. Corning.
\newblock ``{How}'' vs. ``{Why}'' questions in symbiogenesis, and the causal
  role of synergy.
\newblock {\em Biosystems}, 205:104417, 2021.

\bibitem{Doncieux2015}
S.~Doncieux, N.~Bredeche, J.-B. Mouret, and A.E. Eiben.
\newblock Evolutionary robotics: what, why, and where to.
\newblock {\em Frontiers in Robotics and AI}, 2:4, 2015.

\bibitem{emmeche2000levels}
C.~Emmeche, S.~K{\o}ppe, and F.s Stjernfelt.
\newblock Levels, emergence, and three versions of downward causation.
\newblock In P.B. Andersen, C.~Emmeche, N.O. Finnemann, and P.V. Christiansen,
  editors, {\em Downward causation. Minds, bodies and matter}, pages 13--34.
  Aarhus University Press, {\AA}rhus, Denmark, 2000.

\bibitem{eng2019microbial}
A.~Eng and E.~Borenstein.
\newblock Microbial community design: methods, applications, and opportunities.
\newblock {\em Current opinion in biotechnology}, 58:117--128, 2019.

\bibitem{feng2018metagenomic}
G.~Feng, T.~Xie, X.~Wang, J.~Bai, L.~Tang, H.~Zhao, W.~Wei, M.~Wang, and
  Y.~Zhao.
\newblock Metagenomic analysis of microbial community and function involved in
  cd-contaminated soil.
\newblock {\em BMC microbiology}, 18(1):1--13, 2018.

\bibitem{fulda2017natural}
F.C. Fulda.
\newblock Natural agency: The case of bacterial cognition.
\newblock {\em Journal of the American Philosophical Association}, 3(1):69--90,
  2017.

\bibitem{Gibson1966}
J.J. Gibson.
\newblock {\em The senses considered as perceptual systems}.
\newblock Houghton Mifflin, 1966.

\bibitem{gorter2020understanding}
F.A. Gorter, M.~Manhart, and M.~Ackermann.
\newblock Understanding the evolution of interspecies interactions in microbial
  communities.
\newblock {\em Philosophical Transactions of the Royal Society B},
  375(1798):20190256, 2020.

\bibitem{gould1982exaptation}
S.J. Gould and E.S. Vrba.
\newblock Exaptation --- a missing term in the science of form.
\newblock {\em Paleobiology}, pages 4--15, 1982.

\bibitem{guzman2018using}
M.~Guzman-Rodriguez, J.A.K. McDonald, R.~Hyde, E.~Allen-Vercoe, E.C. Claud,
  P.M. Sheth, and E.O. Petrof.
\newblock Using bioreactors to study the effects of drugs on the human
  microbiota.
\newblock {\em Methods}, 149:31--41, 2018.

\bibitem{hale2019robot}
M.~Hale, E.~Buchanan~Berumen, A.~Winfield, J.~Timmis, E.~Hart, A.E. Eiben,
  W.~Li, and A.~Tyrrell.
\newblock The {ARE} robot fabricator: How to (re) produce robots that can
  evolve in the real world.
\newblock In {\em International Society for Artificial Life: ALIFE2019}, pages
  95--102. York, 2019.

\bibitem{Harnad1990}
S.~Harnad.
\newblock The symbol grounding problem.
\newblock {\em Physica D: Nonlinear Phenomena}, 42(1-3):335--346, 1990.

\bibitem{heras2019philosophy}
M.~Heras-Escribano.
\newblock {\em The philosophy of affordances}.
\newblock Springer, 2019.

\bibitem{Hidalgo2014}
J.~Hidalgo, J.~Grilli, S.~Suweis, M.A. Mu\~noz, J.R. Banavar, and A.~Maritan.
\newblock Information-based fitness and the emergence of criticality in living
  systems.
\newblock {\em PNAS}, 111:10095--10100, 2014.

\bibitem{igamberdiev2021mathematics}
A.U. Igamberdiev and J.E. Brenner.
\newblock Mathematics in biological reality: The emergence of natural
  computation in living systems.
\newblock {\em Biosystems}, 204:104395, 2021.

\bibitem{jablonka2002changing}
E.~Jablonka and M.J. Lamb.
\newblock The changing concept of epigenetics.
\newblock {\em Annals of the New York Academy of Sciences}, 981(1):82--96,
  2002.

\bibitem{jaeger_fourth_2022}
J.~Jaeger.
\newblock The {Fourth} {Perspective}: evolution and organismal agency.
\newblock In M.~Mossio, editor, {\em Organization in {Biology}}. Springer,
  forthcoming, Berlin, Germany, 2022.
\newblock (preprint: https://osf.io/2g7fh).

\bibitem{jamone2016}
L.~Jamone, E.~Ugur, A.~Cangelosi, L.~Fadiga, A.~Bernardino, J.~Piater, and
  J.~Santos-Victor.
\newblock Affordances in psychology, neuroscience, and robotics: A survey.
\newblock {\em IEEE Transactions on Cognitive and Developmental Systems},
  10(1):4--25, 2016.

\bibitem{kauffman1969metabolic}
S.A. Kauffman.
\newblock Metabolic stability and epigenesis in randomly constructed genetic
  nets.
\newblock {\em Journal of theoretical biology}, 22(3):437--467, 1969.

\bibitem{kauffman1970articulation}
S.A. Kauffman.
\newblock Articulation of parts explanation in biology and the rational search
  for them.
\newblock In {\em Topics in the Philosophy of Biology}, pages 245--263.
  Springer, 1970.

\bibitem{Kau1993}
S.A. Kauffman.
\newblock {\em The origins of order: Self-Organization and Selection in
  Evolution}.
\newblock Oxford University Press, 1993.

\bibitem{kauffman1996home}
S.A. Kauffman.
\newblock {\em At home in the universe: The search for the laws of
  self-organization and complexity}.
\newblock Oxford university press, 1995.

\bibitem{kauffman2010coevolution}
S.A. Kauffman.
\newblock Co-evolution to the edge of chaos.
\newblock {\em Artificial Life II}, pages 325--370, 2010.

\bibitem{Kauffman2016-humanity}
S.A. Kauffman.
\newblock {\em Humanity in a creative universe}.
\newblock Oxford University Press, 2016.

\bibitem{Kauffman2019beyond}
S.A. Kauffman.
\newblock {\em A world beyond physics: the emergence and evolution of life}.
\newblock Oxford University Press, 2019.

\bibitem{kauffman2021world}
S.A. Kauffman and A.~Roli.
\newblock The world is not a theorem.
\newblock {\em Entropy}, 23(11):1467, 2021.

\bibitem{kauffman2021third}
S.A. Kauffman and A.~Roli.
\newblock The third transition in science: {Beyond} {Newton} and quantum
  mechanics --- {A} statistical mechanics of emergence.
\newblock To appear.
\newblock Preprint available as arXiv:2106.15271.

\bibitem{koza1991genetic}
J.R. Koza.
\newblock Genetic evolution and co-evolution of computer programs.
\newblock {\em Artificial life II}, 10:603--629, 1991.

\bibitem{kull1998semiosis}
K.~Kull.
\newblock On semiosis, umwelt, and semiosphere.
\newblock {\em Semiotica}, 120:299--310, 1998.

\bibitem{kull2015evolution}
K.~Kull.
\newblock Evolution, choice, and scaffolding: Semiosis is changing its own
  building.
\newblock {\em Biosemiotics}, 8(2):223--234, 2015.

\bibitem{longo2013perspectives}
G.~Longo and M.~Mont{\'e}vil.
\newblock {\em Perspectives on organisms: Biological time, symmetries and
  singularities}.
\newblock Springer, 2013.

\bibitem{margulis2008symbiotic}
L.~Margulis.
\newblock {\em Symbiotic planet: a new look at evolution}.
\newblock Basic books, New York, NY, USA, 2008.

\bibitem{metch2018metagenomic}
J.W. Metch, N.D. Burrows, C.J. Murphy, A.~Pruden, and P.J. Vikesland.
\newblock Metagenomic analysis of microbial communities yields insight into
  impacts of nanoparticle design.
\newblock {\em Nature nanotechnology}, 13(3):253--259, 2018.

\bibitem{munoz2018colloquium}
M.A. Munoz.
\newblock Colloquium: Criticality and dynamical scaling in living systems.
\newblock {\em Reviews of Modern Physics}, 90(3):031001, 2018.

\bibitem{noble2012theory}
D.~Noble.
\newblock A theory of biological relativity: no privileged level of causation.
\newblock {\em Interface focus}, 2(1):55--64, 2012.

\bibitem{Nolfi-and-Floreano2000}
S.~Nolfi and D.~Floreano.
\newblock {\em Evolutionary robotics}.
\newblock The MIT Press, Cambridge, MA, 2000.

\bibitem{pascual2020metabolically}
A.~Pascual-Garc{\'\i}a, S.~Bonhoeffer, and T.~Bell.
\newblock Metabolically cohesive microbial consortia and ecosystem functioning.
\newblock {\em Philosophical Transactions of the Royal Society~B},
  375(1798):20190245, 2020.

\bibitem{peil2014emotion}
K.T. Peil.
\newblock Emotion: the self-regulatory sense.
\newblock {\em Global advances in health and medicine}, 3(2):80--108, 2014.

\bibitem{persons2015bristles}
W.S. Persons~IV and P.J. Currie.
\newblock Bristles before down: a new perspective on the functional origin of
  feathers.
\newblock {\em Evolution}, 69(4):857--862, 2015.

\bibitem{pfeifer2006How}
R.~Pfeifer and J.~Bongard.
\newblock {\em How the Body Shapes the Way We Think: A New View of
  Intelligence}.
\newblock MIT Press, Cambridge, MA, 2006.

\bibitem{pollack2003computer}
J.B. Pollack, G.S. Hornby, H.~Lipson, and P.~Funes.
\newblock Computer creativity in the automatic design of robots.
\newblock {\em Leonardo}, 36(2):115--121, 2003.

\bibitem{prosser2020putting}
J.I. Prosser.
\newblock Putting science back into microbial ecology: a question of approach.
\newblock {\em Philosophical Transactions of the Royal Society B},
  375(1798):20190240, 2020.

\bibitem{prosser2020conceptual}
J.I. Prosser and J.B.H. Martiny.
\newblock Conceptual challenges in microbial community ecology.
\newblock {\em Philosophical Transactions of the Royal Society~B},
  375(1798):20190241, 2020.

\bibitem{prum2002evolutionary}
R.O. Prum and A.H. Brush.
\newblock The evolutionary origin and diversification of feathers.
\newblock {\em The Quarterly review of biology}, 77(3):261--295, 2002.

\bibitem{ravikrishnan2018systems}
A.~Ravikrishnan and K.~Raman.
\newblock {\em Systems-level modelling of microbial communities: theory and
  practice}.
\newblock CRC Press, Boca Raton, FL, USA, 2018.

\bibitem{ray_evolution_1992}
T.~S. Ray.
\newblock Evolution and optimization of digital organisms.
\newblock In K.~R. Billingsley, H.~U. Brown, and E.~Derohanes, editors, {\em
  Scientific Excellence in Supercomputing: the 1990 {IBM} Contest Prize
  Papers}, pages 489--531. Baldwin Press, Atlanta, GA, USA, 1992.

\bibitem{riesenfeld2004metagenomics}
C.S. Riesenfeld, P.D. Schloss, and J.~Handelsman.
\newblock Metagenomics: genomic analysis of microbial communities.
\newblock {\em Annual review of genetics}, 38(1):525--552, 2004.

\bibitem{roli2022organisms}
A.~Roli, J.~Jaeger, and S.A. Kauffman.
\newblock How organisms come to know the world: fundamental limits on
  artificial general intelligence.
\newblock {\em Frontiers in Ecology and Evolution}, page 1035, 2022.

\bibitem{roli2020emergence}
A.~Roli and S.A. Kauffman.
\newblock Emergence of organisms.
\newblock {\em Entropy}, 22(10):1163:1--12, 2020.

\bibitem{roli2018dynamical}
A.~Roli, M.~Villani, A.~Filisetti, and R.~Serra.
\newblock Dynamical criticality: overview and open questions.
\newblock {\em Journal of Systems Science and Complexity}, 31(3):647--663,
  2018.

\bibitem{sharov2016evolutionary}
A.A. Sharov.
\newblock Evolutionary biosemiotics and multilevel construction networks.
\newblock {\em Biosemiotics}, 9(3):399--416, 2016.

\bibitem{siggins2012exploring}
A.~Siggins, E.~Gunnigle, and F.e Abram.
\newblock Exploring mixed microbial community functioning: recent advances in
  metaproteomics.
\newblock {\em FEMS microbiology ecology}, 80(2):265--280, 2012.

\bibitem{thorson2017epigenetics}
J.L.M. Thorson, M.~Smithson, D.~Beck, I.~Sadler-Riggleman, E.~Nilsson,
  M.~Dybdahl, and M.K. Skinner.
\newblock Epigenetics and adaptive phenotypic variation between habitats in an
  asexual snail.
\newblock {\em Scientific Reports}, 7(1):1--11, 2017.

\bibitem{Uexkull-foray-correct}
{Uexk{\"u}ll, J.~von}.
\newblock {\em A foray into the worlds of animals and humans: With a theory of
  meaning}.
\newblock University of Minnesota Press, 2010.
\newblock Trad. by J.D. O'Neil. Originally published in German in 1934.

\bibitem{walsh2015organisms}
D.M. Walsh.
\newblock {\em Organisms, agency, and evolution}.
\newblock Cambridge University Press, 2015.

\bibitem{West-Eberhard6543}
M.J. West-Eberhard.
\newblock Developmental plasticity and the origin of species differences.
\newblock {\em Proceedings of the National Academy of Sciences}, 102(suppl.
  1):6543--6549, 2005.

\end{thebibliography}
\end{document}